\documentclass[prl,twocolumn,showpacs,superscriptaddress,floatfix]{revtex4}
\usepackage{graphicx}
\usepackage{epsfig}

\begin{document}
\title{Strongly Resonant Transmission of Electromagnetic Radiation in Periodic Anisotropic Layered Media}

\author{A.A.~Chabanov}
\affiliation{Department of Physics and Astronomy,
University of Texas, San Antonio, TX 78249}

\date{September 9, 2007}

\begin{abstract}
The electromagnetic dispersion in periodic layered media can be tailored and their resonant properties can be considerably improved by utilizing anisotropic materials. Periodic structures with a photonic band edge split into two parts, or so-called split band edge, exhibit superior resonant properties including exceptionally high $Q$-values of transmission resonances and nearly perfect impedance matching at the boundaries, even when the number of unit cells $N$ is not large. A microwave transmission resonance with $Q$$\sim$220 is demonstrated in a periodic stack of form-birefringent layers with $N$=12 realized in a waveguide geometry.
\end{abstract}


\maketitle
The propagation of electromagnetic waves in periodic layered media has been of considerable interest in recent years because of the properties of frequency band gaps and transmission band-edge resonances \cite{Yariv_book,Yeh_book,Scalora, Vitebsky}. When the wave frequency falls within a band gap of periodic layered medium, the electromagnetic energy is nearly totally reflected, and the medium acts as a high-reflectivity reflector for the incident wave. But when the frequency is tuned to the transmission resonance, the wave propagation exhibits a transmittance of nearly unity, long lifetime and significant field amplitude growth within the medium. These phenomena have been utilized in numerous applications including Bragg reflectors, optical delay lines \cite{Scalora_PRE96}, lasing \cite{lase1,lase2}, nonlinear optical devices \cite{Sankley_APL92,Gesuele_JPhysD07}, as well as to increase the sensitivity of embedded sensor and antenna arrays \cite{Volakis}.

The resonance efficiency in slowing down electromagnetic waves and generating high gain depends on how close the transmission resonance is to stationary points of the dispersion relation $\omega(k)$ of the periodic medium, at which the group velocity vanishes, $v_{\rm g}$$\equiv\,$$\partial\omega/\partial k$=0. Here $k$ is the Bloch wave number. In periodic media formed from {\it isotropic} layers of alternating materials with varying refractive index $n$, stationary points occur at band edges which coincide with the origin and boundary of the Brillouin zone, $k$=0 and $\pi/a$, respectively, where $a$ is the period of the layered structure (Fig.~1a). At the transmission resonance, the electromagnetic field inside the medium is nearly a standing wave comprising a pair of counter-propagating Bloch waves. At the resonance closest to the band edge $g$ in Fig.~1a, the wave numbers of the Bloch wave components are given by $k_r^{(\pm)}$=$\pi/a$$\pm$$\pi/Na$, where $N$ is the number of repeated pairs of low/high refractive index material. The resonance occurs in any polarization, since the medium is isotropic. Since the dispersion relation $\omega(k)$ at the {\it regular} band edge (RBE), as in Fig.~1a, can be approximated as a parabola, $\Delta\omega$$\,\sim\,$$(\Delta k)^2$, the resonance frequency $\omega(k_r^{(\pm)})$, is given by $\omega_r$$\,\approx\,$$\omega_g$+$(\omega^{''}_g/2)(\pi/Na)^2$, where $\omega_g$=$\omega(k_g)$ is the RBE frequency and $\omega^{''}_g$=$(\partial^2\omega/\partial k^2)_{k=k_g}$. It can then be shown that the resonant lifetime and field intensity within the medium are proportional to $N^2$, for $N$$\gg$1 \cite{Scalora,Vitebsky}.
\begin{figure}
\includegraphics[width=\columnwidth]{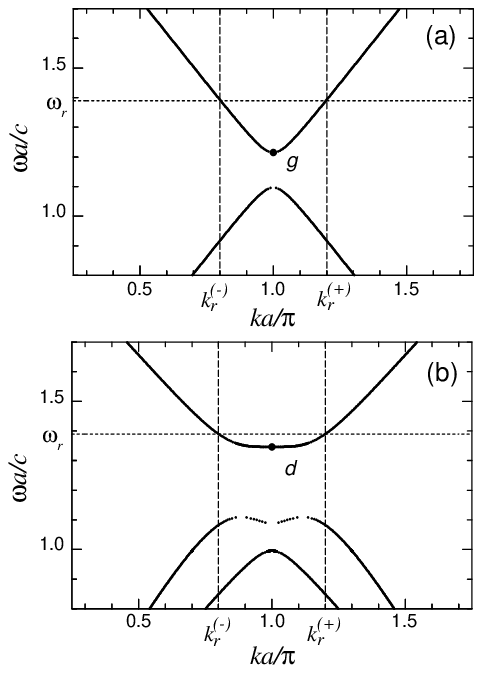}
\caption{Photonic band diagrams featuring (a) Regular Band Edge $g$, (b) Degenerate Band Edge $d$ and Split Band Edge (below the gap). $\omega_r$ indicates the position of the band-edge transmission resonance; $k_r^{(\pm)}$ are the wave numbers of the Bloch waves at the resonance.}
\end{figure}

The resonant properties of periodic layered media can be further improved by utilizing anisotropic materials \cite{Vitebsky}. Particularly, periodic stacks of layers with {\it in-plane} anisotropy misaligned combined with isotropic layers have been shown to exhibit the {\it degenerate} band edge (DBE) \cite{Vitebsky} with a much flatter $\omega(k)$, as compared to RBE. At the DBE, as in Fig.~1b, the dispersion curve is a bi-parabola, $\Delta\omega$$\,\sim\,$$(\Delta k)^4$, and the resonance frequency is $\omega_r$$\,\approx\,$$\omega_d$+$(\omega^{''''}_d/24)(\pi/Na)^4$, where $\omega_d$=$\omega(k_d)$ is the DBE frequency and $\omega^{''''}_d$=$(\partial^4\omega/\partial k^4)_{k=k_d}$. The resonant lifetime and field intensity within the medium are proportional to $N^4$, rather than $N^2$, for $N$$\gg$1. Thus, the DBE layered stack of thickness $L$=$N\!a$ may perform as well as the RBE stack of thickness $N^2a$.

The DBE layered structure is, however, more complex. It includes three layers per unit cell, two anisotropic and one isotropic, and is realized for a single misalignment angle of the anisotropic layers, $\phi$=$\phi_{\rm DBE}$. When the misalignment angle $\phi$ deviates from $\phi_{\rm DBE}$, the DBE splits into two RBEs symmetrical about $k$=$\pi/a$, as below the band gap in Fig.~1b. This band edge configuration will be referred to as the {\it split} band edge (SBE). Note that the SBE may be superior to the DBE in achieving strongly resonant transmission. As the deviation of $\phi$ from $\phi_{\rm DBE}$ is increased, for a given $N$, the RBEs diverging from $k$=$\pi/a$ move closer to and may even coincide with $k_r^{(\pm)}$, leading to vanishing $v_{\rm g}$. This is in contrast to the RBE and DBE layered structures, in which the $k_r^{(\pm)}$ are pushed closer to the band edge by increasing the number of unit cells, $N$.

Here, the resonant transmission properties of the SBE layered structures are studied with computer simulations and microwave measurements. It is shown that strongly resonant transmission near the SBE can always be achieved, for a given $N$, by adjusting the misalignment angle $\phi$. The SBE layered structure exhibits superior resonant properties even for moderate $N$. Specifically, the $Q$ of the SBE resonator may exceed that of a defect-localized state in the middle of the gap of the SBE layered structure. In addition, resonant transmission is nearly unity for any incident polarization, due to a pair of distinct SBE transmission resonances which occur in crossed polarizations. The details of the findings are reported below.

Periodic structures studied in computer simulations included three layers per unit cell, arranged as B-A$_1$-A$_2$. The A-layers were birefringent, with the refractive indices $n_x$=2.87 and $n_y$=1.15 in the plane of layers, and with anisotropy axes of the A$_2$-layers misaligned by angle $\phi$ relative to the A$_1$-layers. The A-layers were of the same thickness, $d_A$=$0.34a$. The B-layers were isotropic, with a refractive index of 4. All the layers were assumed to be lossless.

When $\phi$=$45^0$, the structure produces a DBE above the band gap, as in Fig.~1b. When $\phi$=$30^0$, the DBE is transformed into a SBE shown by a dotted line in Fig.~2a. To achieve a SBE transmission resonance, the resonance frequency $\omega_r$ needs to fit in between the crests and the trough of the SBE in Fig.~2a. This occurs, for example, for $N$=16. In Fig.~2, the calculated transmission through the SBE layered stack of $N$=16 shows a strong resonance of $Q$$\sim$2200 for the incident polarization along $x$-axis of the A$_1$-layers (solid line). This transmission resonance is associated with a pair of counter-propagating Bloch modes with $k_{r1}^{(\pm)}$ (Fig.~2a). The other two modes with $k_{r2}^{(\pm)}$ produce another transmission resonance of $Q$$\sim$120 for the incident polarization along $y$-axis of the A$_1$-layers (dashed line). Thus, the resonant properties of the SBE layered structure are drastically changed by changing polarization of the incident wave. Note that the resonant transmission through the SBE layered stack is nearly unity for any incident polarization, as the transmission resonances occur in crossed polarizations. This is in contrast to the DBE configuration, where only a half of the incident unpolarized light is transmitted, since near the DBE two of the four Bloch modes are propagating and the other two are evanescent.

\begin{figure}[!]
\includegraphics[width=\columnwidth]{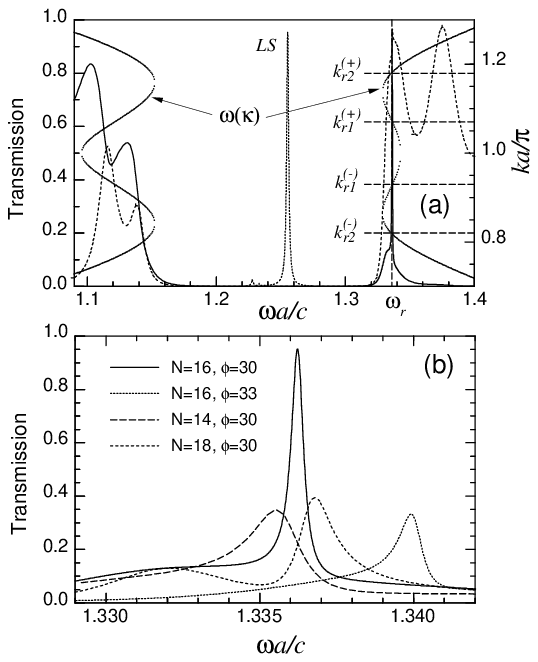}
\caption{(a) Dispersion relation $\omega(k)$ (dotted line, right axis) and transmission spectra (left axis) of the SBE layered stack of $N$=16 and $\phi$=30$^0$. A vertical long-dash line marks the position of SBE transmission resonances occurring in crossed polarizations at $\omega$=$\omega_r$. $k_{r1}^{(\pm)}$ and $k_{r2}^{(\pm)}$ are the respective wave numbers of the two resonances. Transmission spectra are computed for nearly-eigen-polarizations (solid and short-dash lines). {\it LS} (dotted line) denotes transmission via a midgap localized state due to a twist defect in the structure. (b) Zoomed-in transmission resonance computed for various $N$ and $\phi$.}
\end{figure}

The SBE layered structure is not only far superior to the DBE structure as a resonator, but is also superior to the SBE structure with a defect-localized state in the gap. Localized electromagnetic states due to defects in periodic structures are known to have exceptionally high $Q$'s, although enhancement of gain is limited to a small volume at the defect, as compared to gain volume of the transmission resonance. Localized states in the band gap of the SBE layered structure can be created by introducing a twist defect, a quarter-wavelength separation, or a combination of two in the center of the structure \cite{Hodg,Kopp}. None of these, however, produces $Q$ as high as that of the SBE transmission resonance. For example, a twist of 45$^0$ in the center of the $N$=16 structure creates a pair of localized states in the middle of the gap. Transmission via one of the localized states with $Q$$\sim$700 is plotted in Fig.~2a ({\it LS}, dotted line).

When the number of unit cells is changed from $N$=16, while the $\phi$=30$^0$ is kept unchanged, the high-$Q$ resonance disappears, notwithstanding the resonant frequency falls between and even closer to the stationary points of the SBE (Fig.~2b). Changing of $\phi$ from 30$^0$, for $N$=16, has a similar affect on the resonance. Nevertheless, a high-$Q$ transmission resonance, as seen in Fig.~2, can always be achieved for practically any $N$ by adjusting the misalignment angle $\phi$.

The occurrence of high-$Q$ transmission resonance in SBE layered structures has been tested in microwave experiment. Dielectric layers matching those studied in the computer simulations were fabricated of low-loss microwave ceramics in the form of disks of diameter 20.64 mm (Fig.~3a). A-layers of thickness 4.14 mm were constructed of six parallel flat ceramic pieces of $n$=5.48 and loss tangent $\delta$=5$\times$$10^{-4}$, held in places with a thin plastic ring. The thickness and separation of ceramic pieces were made to induce $n$=2.87 and 1.15, for polarizations parallel and perpendicular to the pieces, respectively \cite{Wolf}. To align the respective layers of the A$_1$ and A$_2$ and to control the misalignment angle $\phi$, indexing pins on the side of plastic rings were used. Isotropic B-layers were solid disks of thickness 1.25 mm, $n$=4 and $\delta$=$10^{-4}$.

A loop of Teflon rods was employed to deliver microwave radiation to and from the periodic layered samples. Two horns, set to produce radiation of the same polarization, were used to launch and detect the microwave radiation. Field spectra in samples with $N$ ranging from 8 to 17 were obtained with use of a Hewlett-Packard N5230A vector network analyzer. Once a band-edge transmission resonance was detected for a given $N$, the misalignment angle $\phi$ and sample orientation relative to the incident polarization were adjusted to achieve the highest $Q$. A typical spectrum of polarized transmission measured in a sample of $N$=12 and $\phi$=30$\pm$1$^0$ is shown in Fig.~3b (solid line). A peak with $Q$$\sim$220 in the transmission spectrum indicates a resonance  at $\nu$=7.34 GHz. The resonant character of transmission is further indicated by a sharp peak in the spectrum of group delay (dotted line). The group delay, as measured by the network analyzer, is given by $\tau$=$d\varphi/d\omega$, where $\varphi$ is the phase of the field and $\omega$ is the angular frequency. A peak value of $\tau$=7.1 ns gives the microwave leakage time from the sample of $L$=11.5 cm. A second peak in the group delay spectrum at $\nu$=7.29 GHz marks the position of the band edge.

\begin{figure}[t!]
\includegraphics[width=\columnwidth]{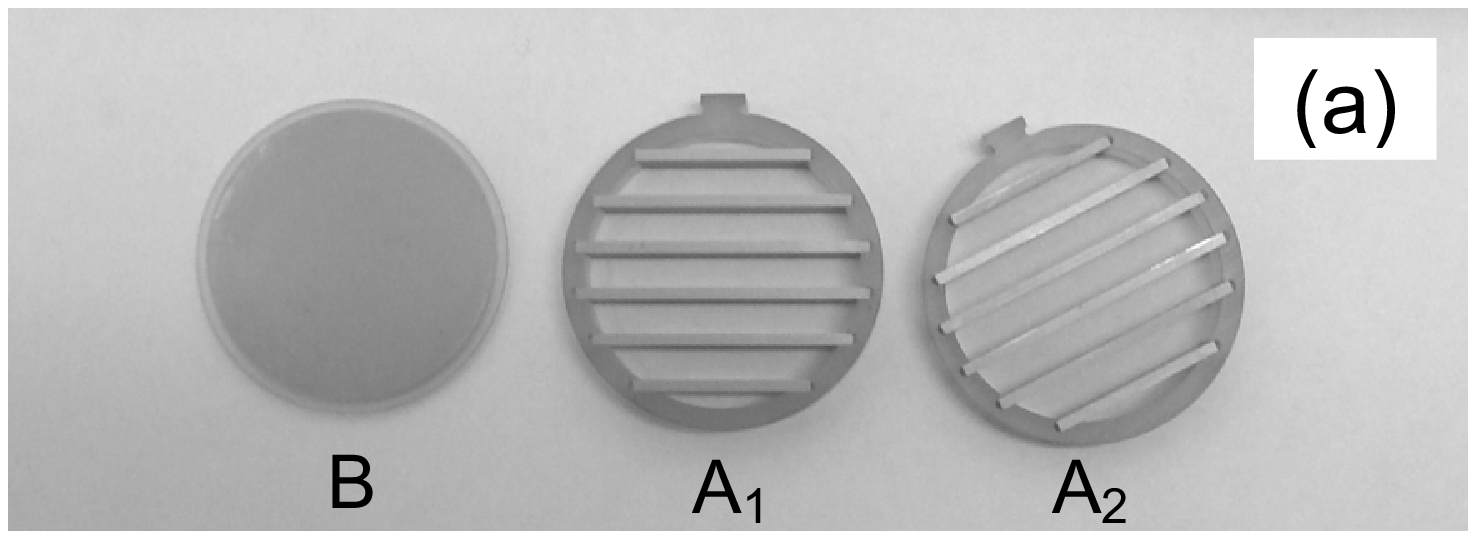}
\includegraphics[width=\columnwidth]{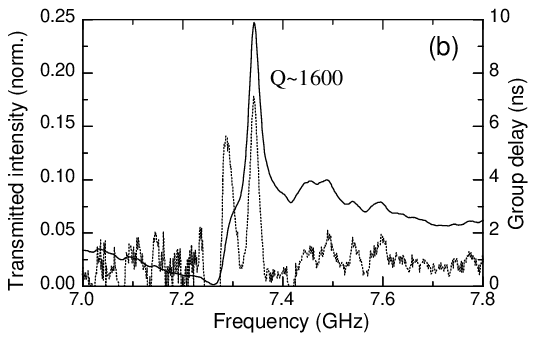}
\caption{(a) Picture of the fabricated A- and B-layers of the periodic layered samples used in the microwave experiment. (b) Spectra of polarized transmitted intensity and group delay measured in the sample of $N$=12 and $\phi$=30$\pm$1$^0$. The intensity spectrum is normalized relative to that measured without sample. Peaks in the intensity and group delay spectra indicate a SBE transmission resonance of $Q$$\sim$220 at $\nu$=7.34 GHz.}
\end{figure}

The microwave measurements thus confirm that the transmission resonance of high $Q$ can be achieved in the SBE layered structure, even when $N$ is not large. One can, however, notice that the measured spectra are slightly blue-shifted, as compared to calculations, and that the highest $Q$ in the sample of $\phi$=30$^0$ is achieved for $N$=12, rather than 16. In addition, a peak value of the normalized transmitted intensity in Fig.~3b is only 0.25. The discrepancies between the calculated and measured results may be attributed to waveguide geometry of the samples, which was not included in the calculations. The samples studied supported a few guided modes at the resonant frequency, whereas the Teflon rods supported a single mode. Thus, the transmitted intensity could not be properly normalized. Reduced values of the measured transmission are also due to an arbitrary polarization of transmitted wave relative to the detector. In addition, the presence of higher-order guided modes in the sample may explain moderate $Q$-values of transmission resonances, as well as a relatively high level of transmission at frequencies above the transmission resonance in Fig.~3b.

In conclusion, transmission resonances of exceptionally high $Q$'s were achieved in periodic anisotropic layered media near the SBE, even when the number of unit cells was not large, which is of critical importance to the fabrication of high-quality photonic structures. The SBE resonators exhibit a nearly perfect impedance matching at the boundaries. The resonant transmission is nearly unity for any incident polarization, while the resonance bandwidth depends on the wave polarization. Resonant transmission near the SBE in the microwave spectral range was demonstrated in periodic stacks of form-birefringent layers realized in waveguide geometry. A similar approach can be used to extend the SBE layered structures to the visible range, in which naturally occurring dielectric anisotropy is too weak.

The author thanks A.Z.~Genack for critical reading the manuscript, V.~Kopp, A.~Figotin and I.~Vitebskiy for valuable discussions,  and P.~Kruger for help in constructing the experimental samples. The calculations were performed with the Light Propagation Simulation Program kindly provided by V.~Kopp. This work was supported by the AFOSR DURIP Grant No.~FA9550-06-1-0389 and UTSA Faculty Research Award.

\end{document}